# Steady state current transfer and scattering theory

Vered Ben-Moshe[1], Dhurba Rai,[1] Spiros Skourtis[2] and Abraham Nitzan[1]

[1] School of Chemistry, Tel Aviv University, Tel Aviv, 69978, Israel
[2] Department of Physics, University of Nicosia, Nicosia, Cyprus

## Abstract

The correspondence between the steady state theory of current transfer and scattering theory in a system of coupled tight-binding models of 1-dimensional wires is explored. For weak interwire coupling both calculations give nearly identical results, except at singular points associated with band edges. The effect of decoherence in each of these models is studied using a generalization of the Liouville-von Neuman equation suitable for steady-state situations. An example of a single impurity model is studied in details, leading to a lattice model of scattering off target that affects both potential scattering and decoherence. For an impurity level lying inside the energy band, the transmission coefficient diminishes with increasing dephasing rate, while the opposite holds for impurity energy outside the band. The efficiency of current transfer in the coupled wire system decreases with increasing dephasing.

## 1. Introduction

In a recent paper[1] we have introduced *Current transfer* as a charge transfer transition chracterized by relocation of both charge and its momentum. In that work[1] current transfer was analyzed in the time domain and was proposed to be the mechanism behind recent observations[2-3] that indicate that photo-electron transfer induced by circularly polarized light through helical molecular bridges depends on the relative handedness of the bridge helicity and on the optical circular polarization. More recently[4] we have analyzed current transfer in steady state situations, where the system response to an imposed current in one of its segment is of interest. While this problem is mathematically well defined and may correspond, at least as an approximation, to situations of physical interest, some of its characteristics may appear unphysical. For example, the steady state current consistent with a given current imposed on part of a system is not subjected to any conservation law and may attain values larger than the imposed current.[4]

A simple example is shown in Fig. 1, which depicts two infinite tight binding wires D ("donor") and A ("acceptor") characterized by lattice constant $a$, site energies $\varepsilon_D, \varepsilon_A$ and nearest-neighbor couplings $\beta_D, \beta_A$, locally coupled to each other by the interaction $V$ that couples a finite number $N_{DA}$ of close proximity sites on the two wires. The Hamiltonian is $\hat{H} = \hat{H}_D + \hat{H}_A + \hat{V}_{DA}$, where (see Fig. 1)

$$\hat{H}_K = \sum_{j \in K} \varepsilon_K |j\rangle\langle j| + \sum_{j \in K} \beta_K |j\rangle\langle j+1| \; ; \quad K = D, A \tag{1}$$

and

$$\hat{V}_{DA} = \sum_{j_D, j_A}^{N_{DA}} V |j_D\rangle\langle j_A| + \text{h.c} \tag{2}$$

A Bloch wavefunction of wavevector $k$ carrying current $J_D = -(2\beta_D/\hbar)\sin(ka)$ is imposed on the wire D

$$\Psi_D(t) \sim e^{-i(E/\hbar)t} \sum_{j_D=1}^{N_D} e^{i(j_D-1)ka} |j_D\rangle \; ; \quad E = E_D + 2\beta_D \cos(ka) \tag{3}$$

and the current on A consistent with this "boundary condition" is evaluated. To this end, the steady state wavefunction on wire A is written in the form

$$\Psi_A(t) = \sum_{j \in A} C_j(t)|j\rangle = e^{-i(E/\hbar)t} \sum_{j \in A} \bar{C}_j |j\rangle \tag{4}$$

where $\bar{C}_j$ are time independent coefficients that satisfy

$$(E - E_j)\bar{C}_j - \sum_k V_{jk}\bar{C}_k = 0 \tag{5}$$

in which $k$ goes over all sites coupled to $j$, with $V_{jk} = \beta_A$ when $k$ is on A, and $V_{jk} = V$ when $k$ is on D. Eq. (5) constitutes an infinite set of equations for the coefficients $\bar{C}_j$, $j \in A$, that contain inhomogeneous terms with $\bar{C}_k, k \in D$. Since the latter are given (Eq. (3)), this provides an expression for the current between any two sites on wire A

$$J_{A(j-1 \to j)} = \frac{2\beta_A}{\hbar} \text{Im}(\bar{C}_{j-1}\bar{C}_j^*) \tag{6}$$

in terms of that imposed on wire D.[4] The solution is facilitated by truncating the infinite set of equations (5) beyond the two wires interaction region, using the known surface self energy of a one-dimensional nearest-neighbor tight-binding lattice. This procedure[4] is reproduced below.

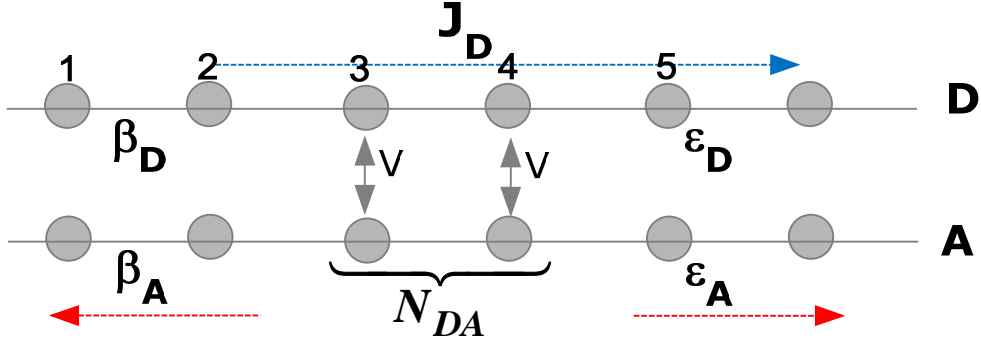

Fig. 1. The current transfer problem in a tight-binding wire system. Wires D and A are coupled to each other at $N_{DA}$ positions. Wire D is restricted to hold a constant Bloch wave of energy $E$ that carries a current $J_D$. The objective is to calculate the current induced on the wire A.

The procedure described above is used to evaluate the steady state current on wire A induced by the imposed Bloch wavefunction on D. In particular, the difference between the currents going to the right and to the left of the interaction region on A is a manifestation of a current transfer property: a charge transfer detector placed on A on the right of the interaction region will be sensitive to the direction of the current on D.[1,4]

The dynamics imposed by a given driving current on the wire D can be contrasted with the more familiar scattering process described by Fig. 2. This process is characterized by four channels, a, b, c, d. It is driven by an incoming wave in channel a, which induces four outgoing waves in channels a-d (the outgoing wave in channel a is the reflected wave). Current conservation now strictly applies: the incoming source flux should be equal to the sum of all outgoing fluxes.

While Figure 2 represents a familiar scattering problem, the process described by Figure 1 is less obvious from the physical point of view. Indeed, the boundary condition that restricts the wavefunction in the D wire to be a Bloch state of given energy and wavevector can be realized only approximately as a strong driving-weak scattering limit. Still, it is a mathematically well defined problem, simpler than the corresponding scattering problem, which provides a reasonable approximation in many situations. In the present paper we compare the two problems and the processes they describe. The solution of the current transfer problem exemplified by Fig. 1 was presented in Ref. [4]. In Section 2 we describe a procedure for solving the corresponding scattering problem using a similar steady-state approach, and compare the two processes, focusing on several prototypical models. Section 3 describes an approximate solution to the scattering problem in the presence of dephasing, again comparing simple model results with the exact solutions of corresponding current transfer problems. Section 4 concludes.

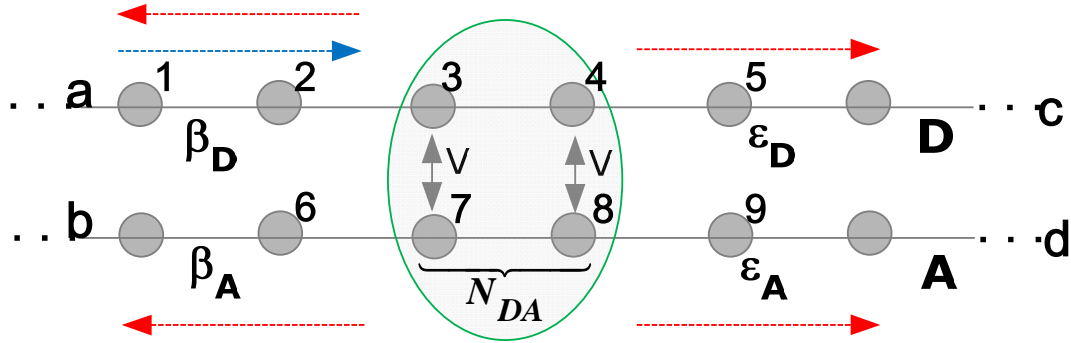

Fig. 2. The scattering problem equivalent to the current transfer problem of Fig. 1. An incident bloch particle in channel a scatters from an 'impurity center' (encircled) into four outgoing waves in channels a-d, including the reflected wave in channel a. The impurity center comprises $N_{DA}$ pairs of sites that link between the wires. Here $N_{DA} = 2$.

## 2. The scattering formalism in coupled wires systems

The method of solution of this scattering problem may be illustrated by the simpler scattering problem of Fig. 3. Consider the steady state driven by the incoming Bloch wave of energy $E$. This waves scatters from the impurity center at site 3, generating the transmitted and reflected waves $J_T$ and $J_R$, respectively. We take all site energies to be $\alpha$, except the energy impurity site $\varepsilon_3$, and the (assumed real) nearest neighbor coupling is denoted $\beta$. At steady state, the coefficient of the wavefuncion in the site representation,

$$\Psi(t) = e^{-i(E/\hbar)t} \sum_j \bar{C}_j |j\rangle \qquad (7)$$

(as in Section 1, the coefficients $\bar{C}_j$ are time independent) satisfy equations analogous to Eq. (5),

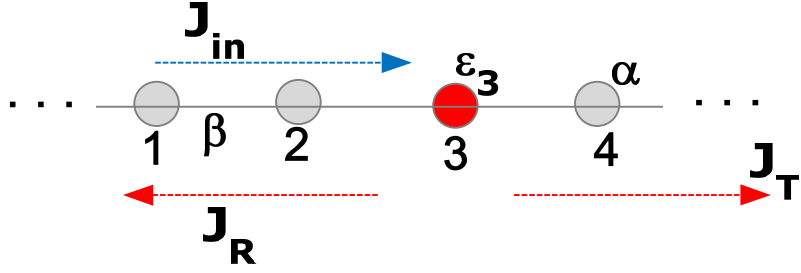

Fig. 3. A simple model demonstrating the scattering calculation described in the text

$$(E-\alpha)\bar{C}_2 - \beta(\bar{C}_1 + \bar{C}_3) = 0 \qquad (8a)$$

$$(E-\varepsilon_3)\bar{C}_3 - \beta(\bar{C}_2 + \bar{C}_4) = 0 \qquad (8b)$$

$$(E-\alpha-\Sigma(E))\bar{C}_4 - \beta\bar{C}_3 = 0 \qquad (8c)$$

In fact, we can truncate this set of equation already at site 3:

$$(E-\alpha)\bar{C}_2 - \beta(\bar{C}_1 + \bar{C}_3) = 0 \qquad (9a)$$

$$(E-\varepsilon_3-\Sigma(E))\bar{C}_3 - \beta\bar{C}_2 = 0 \qquad (9b)$$

Indeed, solving (8c) for $\bar{C}_4$,

$$\bar{C}_4 = -\frac{\beta\bar{C}_3}{\alpha-E+\Sigma(E)} \qquad (10)$$

inserting the solution to (8b) and comparing the resulting equation to (9b) yields

$$\Sigma(E) = \frac{(E-\alpha)-\sqrt{(E-\alpha)^2-4\beta^2}}{2} \equiv \Lambda(E)-(1/2)i\Gamma(E) \qquad (11)$$

with $\Lambda$ real and $\Gamma$ real and positive. Note that $\Gamma(E)=0$ unless $E$ is within the energy band defined in Eq. (3), i.e. $\alpha-2|\beta|<E<\alpha+2|\beta|$. Stability considerations dictate the choice of the minus sign in front of the square root.

Another consistency check is to note that Eq. (10) implies that the steady state current from site 3 to 4 (cf. Eq. (6)) is

$$J_{3\to4} = \frac{2\beta}{\hbar}\operatorname{Im}(\bar{C}_4^*\bar{C}_3) = -\frac{2}{\hbar}|\bar{C}_4|^2\operatorname{Im}(\Sigma(E)) = \frac{\Gamma(E)}{\hbar}|\bar{C}_4|^2 \qquad (12)$$

i.e. the flux to the right out of site 4.

Eqs. (9) can be solved to yield

$$\bar{C}_2 = K_2(E)\bar{C}_1 \; ; \quad \bar{C}_3 = K_3(E)\bar{C}_1 \tag{13}$$

$$K_2(E) = \frac{\beta}{E - \alpha - \dfrac{\beta^2}{E - \varepsilon_3 - \Sigma(E)}} \tag{14a}$$

$$K_3(E) = \frac{\beta}{E - \varepsilon_3 - \Sigma(E)} K_2(E) \tag{14b}$$

Up to this point, the solution representing the coefficients in Eq. (7) in terms of the 'driving' term $\bar{C}_1$ is analogous to that of the current transfer problem. However now we seek a solution which to the left of the scattering region is represented by a linear combination of incoming and reflected waves of energy $E$, and on the right of that region, by a transmitted wave. Setting the origin on site 1 and writing $\bar{C}_1$ as a sum of incoming and reflected amplitudes,

$$\bar{C}_1 = A + B \tag{15}$$

it follows that

$$\bar{C}_2 = Ae^{ika} + Be^{-ika} \tag{16}$$

where, (c.f. Eq. (3)),

$$ka = \pm \arccos\left(\frac{E - \alpha}{2\beta}\right) \tag{17}$$

Equations (15) and (16) imply that the *net* current on the left side of the scattering center is[1]

$$J_{1\to 2} = \frac{2\beta}{\hbar} \text{Im}(\bar{C}_2^* \bar{C}_1) = -\frac{2\beta}{\hbar}(|A|^2 - |B|^2)\sin(ka) = J_{in} - J_R \tag{18}$$

where $J_{in}$ and $J_R$ are the incident and reflected currents, respectively. Also, as required by continuity, it is easy to show (see Appendix A) that

$$J_{2\to 3} = \frac{2\beta}{\hbar} \text{Im}(\bar{C}_3^* \bar{C}_2) = J_{1\to 2} \tag{19}$$

and

$$J_{3\to \text{right}} = \frac{\Gamma(E)}{\hbar}|\bar{C}_3|^2 = J_{1\to 2} \tag{20}$$

---

[1] Note that if our tight binding model is a finite difference representation to a free particle motion, then $\beta < 0$.

where $\Gamma(E) = -2\,\text{Im}(\Sigma(E))$. Eq. (20) represents the transmitted current. To find the incident and reflected currents we use (15) and (16) in Eq. (14a) in order to express the reflected amplitude in terms of the incident amplitude

$$B = -\frac{K_2 - e^{ika}}{K_2 - e^{-ika}} A \tag{21}$$

Similarly, the transmitted amplitude is obtained in the form

$$\bar{C}_3 = K_3 \bar{C}_1 = K_3 \frac{2i \sin(ka)}{K_2 - e^{-ika}} A \tag{22}$$

The incident, transmitted and reflected currents are now given by:

$$J_{in} = \frac{\Gamma(E)}{\hbar}|A|^2 \;;\; J_T = \frac{\Gamma(E)}{\hbar}|C_3|^2 \;;\; J_R = \frac{\Gamma(E)}{\hbar}|B|^2 \tag{23}$$

Consistency with Eq. (18) is implied by[4]

$$\Gamma(E) = 2|\beta \sin(ka)| \;;\qquad \beta \sin(ka) < 0 \tag{24}$$

Finally, the transmission and reflection coefficients, $T(E)$ and $\mathcal{R}(E)$ are given by

$$T(E) = \left|\frac{\bar{C}_3}{A}\right|^2 = \left|\frac{2K_3 \sin(ka)}{K_2 - e^{-ika}}\right|^2 \tag{25}$$

$$\mathcal{R}(E) = \left|\frac{B}{A}\right|^2 = \left|\frac{K_2 - e^{ika}}{K_2 - e^{-ika}}\right|^2 \tag{26}$$

and can be shown to satisfy the conservation condition $T(E) + \mathcal{R}(E) = 1$.

The above example makes it clear how the solution to the scattering problem is obtained as an extension of the procedure for solving the current transfer problem. In both we look for a solution to the Schrödinger equation in the form (7), under some given 'boundary conditions'. In the current transfer problem, eg. Fig. 1, the wavefunction on the upper (driver) wire is known, and in particular the driving wavefunction on sites 3 and 4 is given in the form

$$\Psi(t) = e^{-iEt/\hbar}\left(...\bar{C}_3|3\rangle + \bar{C}_4|4\rangle...\right) \tag{27}$$

with $\bar{C}_4 = \bar{C}_3 e^{ika}$. The other (known) coefficients on the upper wire are irrelevant for this example where only sites 3 and 4 on the driving wire D are connected to the driven wire A. Given these coupling and driving model, the coefficients $\{C_j\}$ of the A wire can be computed as described in Section 1 and Ref. [4], yielding the induced left and right currents on this wire by using Eq. (6). In the corresponding scattering problem, with incoming channel on the left side of the scattering center on the D wire, the driving character is assigned to site 1, i.e. a steady state solution to the time dependent Schrödinger equation for both wires is sought, subjected to the condition $\Psi(t) = ... + \bar{C}_1 e^{-iEt/\hbar} |1\rangle + ...$. This solution relates all the coefficients $C_j = \bar{C}_j e^{-iEt/\hbar}$ in (4) to the driving amplitude $\bar{C}_1$, and directly yields the transmitted currents in all channels, e.g. $J_{D,\text{right}} = (\Gamma_D/\hbar)|\bar{C}_5|^2$ in terms of $|\bar{C}_1|^2$. Expressing these results in terms of the more relevant incident intensity is achieved using the procedure demonstrated in Eqs. (15)-(23)

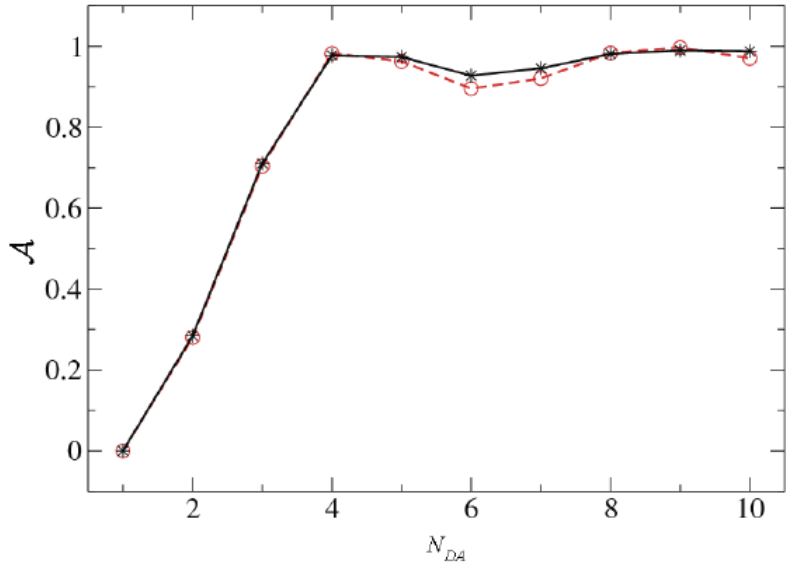

Fig. 5. The current asymmetry factor $\mathcal{A}$ displayed against the number of links, $N_{DA}$ connecting the D and A wires in the DA system (Fig. 2). Parameters are $E_D = E_A = 0$, $\beta_D = \beta_A = 0.1$, $V_{DA} = 0.01$ and the injection energy is $E = -0.15$. Full line (black) – calculation based the scattering model (Fig. 2). Dashed line (red) – calculation based on the current transfer model of Fig. 1.

Figures 5, 7 and 8 compare the results obtained for the current transfer and the scattering calculations using models (Fig. 1 and Fig. 2, respectively) characterized by similar parameters. Fig. 5 shows the current asymmetry factor,

$$\mathcal{A} = \frac{J_A^{left} - J_A^{right}}{J_A^{left} + J_A^{right}} \qquad (28)$$

(where $J_A^{left}$ and $J_A^{right}$ are the steady currents in the A wire to the left and the right of the interaction region, respectively) displayed as a function of the number of links, $N_{DA}$, between the two wires. Fig. 7 shows the dependence of $\mathcal{A}_1$ on the relative band alignments of the two wires, varied by moving the band centers $E_A$ and $E_D$ relative to each other (see Fig. 6). For the chosen parameters both figures show excellent agreement between the two calculations, which deviate from each other only near the band-edge as seen in Fig. 8. The band edge singularity that characterized the current-transfer calculation[4] is absent in the scattering calculation as required by the current conserving nature of the latter.

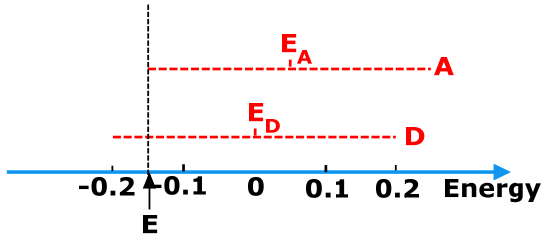

Fig. 6. The energy bands of the D and A wires (The zero order site energies, $E_D$ and $E_A$ correspond to the mid-band energies. For $\beta = 0.1$ the bandwidth is 0.4. For $E_A = 0.05$ an injection energy -0.15 on the D wire corresponds to the band edge on the A wire.

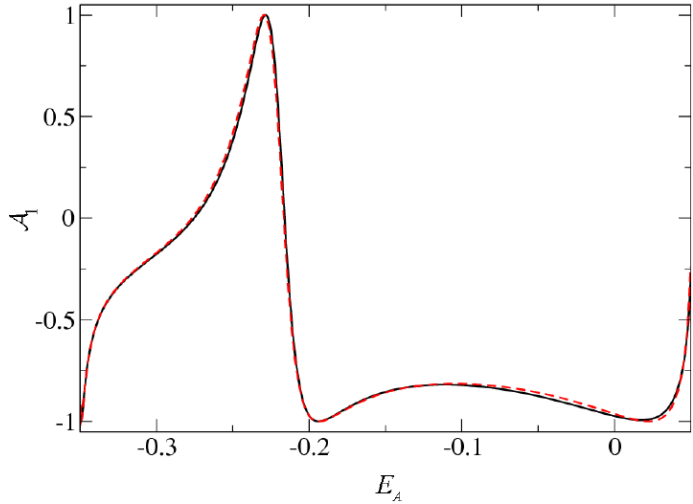

Fig. 7. Current asymmetry factor $\mathcal{A}_1$ displayed as a function of the center $E_A$ of the A-wire band (see Fig. 6), for $E_D = 0$, using $\beta_D = \beta_A = 0.1$, $V = 0.1$ and $N_{DA} = 5$. The injection energy is $E = -0.15$, implying that current can be transmitted to the A wire in for $E_A$ in the range $-0.35 < E_A < 0.05$. Full line (black) – calculation based the scattering model (Fig. 2). Dashed line (red) – calculation based on the current transfer model of Fig. 1.

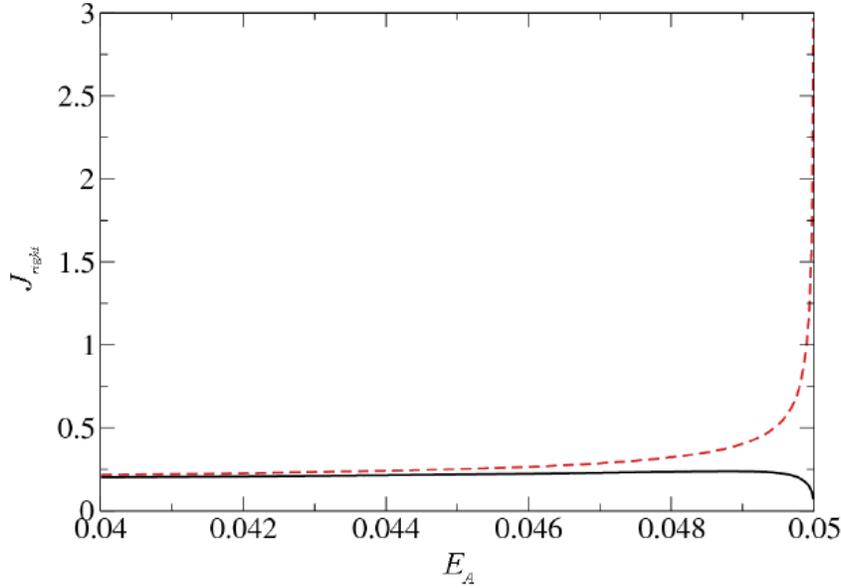

Figure 8. The right-going transmitted current in the A-wire, displayed against the A band center $E_A$ near the transmission threshold $E_A = 0.05$. Note the singularity at $E_A = 0.05$ in the transmitted current calculated from the current transfer model of Fig. 1 that should be contrasted with the regular behavior of the scattering model of Fig. 2. Parameters and line designations are as in Fig. 7.

In Ref. [4] we have argued that current transfer is a coherent phenomenon, resulting from interference between several transport paths, and have studied the effect of dephasing on this process. Next we turn to a similar examination for the equivalent scattering process.

## 3. Scattering in the presence of dephasing

As discussed in Ref. [4], the symmetry breaking in the flux induced in wire A is a manifestation of interference between transport paths. It is therefore sensitive to the way by which the wires link to each other and to the position dependent phase of the carrier wavefunction, in particular at the links positions. The transfer of directionality, is therefore sensitive to dephasing. In Ref. [4] we have examined this issue in the framework of the current transfer problem exemplified in Fig. 1. Here we provide an approximate solution to this problem for the steady-state scattering problem exemplified by Fig. 2. To this end we recast our steady state approach in the Liouville equation framework, following the procedures of Refs. [5] and [4].

While the methodology is general, it is convenient to describe our approach to this problem in terms of the simpler scattering model of Fig. 3, which describes the scattering of a Bloch wave on a 1-dimensional tight-binding lattice from a single impurity site. In the absence of dephasing, the equation of motion for the density matrix elements in the local (site) representation, $\rho_{ij}(t) = C_i(t)C_j^*(t)$, can be obtained in a straightforward way from those for the corresponding coefficients

$$\dot{C}_j = -i\varepsilon_j C_j - i\sum_k V_{jk} C_j, \qquad (29)$$

using $\dot{\rho}_{jk} = C_j \dot{C}_k^* + \dot{C}_j C_k^*$. Here we use the specific site designations for the on-site energies and intersite interactions, $\varepsilon_i$ and $V_{ij}$, that for a uniform lattice were denoted above $\alpha$ and $\beta$, respectively. The effect of dephasing is included in these equations of motion by adding damping terms to the evolution of non-diagonal elements of the density matrix, i.e.,

$$\dot{\rho}_{ij} \to \dot{\rho}_{ij} - \frac{1}{2}(\gamma_i + \gamma_j)(1 - \delta_{ij})\rho_{ij}. \qquad (30)$$

The corresponding steady state equations, $\dot{\rho}_{jk} = 0$ are similar to those used in Refs. [5] and [4].

Focusing on the scattering model of Fig. 3, we start by recasting the infinite set of equations (29) to describe a steady state driven by an oscillating amplitude at site 1, using the known self energy of a particle moving on a 1-dimensional tight binding lattice to represent the dynamics on the finite region of interest. The steady state equations for the amplitudes $\bar{C}_j = C_j \exp(iEt/\hbar)$ then become[4]

$$\begin{aligned}
&\bar{C}_1 = \text{constant} \\
&(E - \varepsilon_2)\bar{C}_2 - V_{21}\bar{C}_1 - V_{23}\bar{C}_3 = 0 \\
&(E - \varepsilon_3)\bar{C}_3 - V_{32}\bar{C}_2 - V_{34}\bar{C}_4 = 0 \\
&(E - \varepsilon_4)\bar{C}_4 - V_{43}\bar{C}_3 - \Sigma_4(E)\bar{C}_4 = 0
\end{aligned} \qquad (31)$$

where

$$\Sigma_4(E) = \frac{E - \alpha - \sqrt{(E-\alpha)^2 - 4\beta^2}}{2} \equiv \Lambda(E) - \frac{i}{2}\Gamma(E) \qquad (32)$$

($\Lambda$ and $\Gamma$ real) is the self energy of site 4 associated with the infinite lattice to its right. The form (32) implicitly assumes that the infinite chain to the right of site 4 is uniform, with equal site energies $\alpha = E_4, E_5...$ and nearest neighbor couplings $\beta = V_{45} = V_{56} = ...$ .

The corresponding steady state equations for $\rho$ should represent all steady state elements $\rho_{jk}$ in terms of the driving site population $\rho_{11}$. In what follows we consider the situation where dephasing originates from dynamical processes on the scattering site 3 only, and take $\gamma_j = \gamma \delta_{j3}$. It is convenient to write the resulting equations in two groups. Those representing $\rho_{j1}$ and $\rho_{1j}$ in terms of $\rho_{11}$ are given by

$$\begin{pmatrix} (E-\varepsilon_2) & -V_{23} & 0 \\ -V_{32} & (E-\varepsilon_3+\tfrac{1}{2}i\gamma) & -V_{34} \\ 0 & -V_{43} & (E-\tilde{\varepsilon}_4+(i/2)\Gamma_4) \end{pmatrix} \begin{pmatrix} \rho_{21} \\ \rho_{31} \\ \rho_{41} \end{pmatrix} = \begin{pmatrix} V_{21}\rho_{11} \\ 0 \\ 0 \end{pmatrix}, \quad (33)$$

(and $\rho_{1j} = \rho_{j1}^*$) where $\tilde{\varepsilon}_4 = \varepsilon_4 + \Lambda_4(E)$, and those expressing $\rho_{jk}$ ($j$ and $k \neq 1$) in terms of $\rho_{j1}$, $\rho_{1j}$ take the form

$$\begin{pmatrix} \varepsilon_{22} & V_{23} & 0 & -V_{23} & 0 & 0 & 0 & 0 & 0 \\ V_{32} & \varepsilon_{32}+\tfrac{1}{2}i\gamma & V_{34} & 0 & -V_{23} & 0 & 0 & 0 & 0 \\ 0 & V_{43} & \varepsilon_{42} & 0 & 0 & -V_{23} & 0 & 0 & 0 \\ -V_{32} & 0 & 0 & \varepsilon_{23}+\tfrac{1}{2}i\gamma & V_{23} & 0 & -V_{34} & 0 & 0 \\ 0 & -V_{32} & 0 & V_{23} & \varepsilon_{33} & V_{34} & 0 & -V_{34} & 0 \\ 0 & 0 & -V_{32} & 0 & V_{43} & \varepsilon_{43}+\tfrac{1}{2}i\gamma & 0 & 0 & -V_{34} \\ 0 & 0 & 0 & -V_{43} & 0 & 0 & \varepsilon_{24} & V_{23} & 0 \\ 0 & 0 & 0 & 0 & -V_{43} & 0 & V_{32} & \varepsilon_{34}+\tfrac{1}{2}i\gamma & V_{34} \\ 0 & 0 & 0 & 0 & 0 & -V_{43} & 0 & V_{43} & \varepsilon_{44} \end{pmatrix} \begin{pmatrix} \rho_{22} \\ \rho_{23} \\ \rho_{24} \\ \rho_{32} \\ \rho_{33} \\ \rho_{34} \\ \rho_{42} \\ \rho_{43} \\ \rho_{44} \end{pmatrix} = V_{21} \begin{pmatrix} \rho_{12}-\rho_{21} \\ \rho_{13} \\ \rho_{14} \\ -\rho_{31} \\ 0 \\ 0 \\ -\rho_{41} \\ 0 \\ 0 \end{pmatrix}$$

(34)

where $\varepsilon_{ij} = \varepsilon_i^* - \varepsilon_j$, $\varepsilon_4 \equiv \tilde{\varepsilon}_4 - (i/2)\Gamma_4$, so that $\varepsilon_{44} = i\Gamma_4(E)$. Note that while Eqs. (34) are derived from the standard Liouville equation, Eq. (33) is a modified form that expresses the driving condition.

Eqs. (33) and (34) can be solved to yield all density matrix elements in terms of $\rho_{11}$,

$$\rho_{ij} = K_{ij}\rho_{11}, \quad (35)$$

where the (in general complex) numbers $K_{ij}$ are obtained from the inverse matrices, and where the diagonal terms $K_{jj}$ are real. In analogy to Eqs. (15) and (16), what we need is to express the density matrix elements in terms of the incident amplitude $A$ (or intensity $|A|^2$). Eqs. (15) and

(16) have however to be modified because the reflected amplitude now assumes a random phase component because of the imposed dephasing on site 3. This is expressed by taking

$$\overline{C}_1 = A + B \, e^{i\varphi} \tag{36a}$$

$$\overline{C}_2 = A e^{ika} + B \, e^{-ika + i\varphi} \tag{36b}$$

(A can be taken real without loss of generality) so that $\rho_{11} = |A|^2 + |B|^2 + A \left\langle \left( B \, e^{i\varphi} + B^* \, e^{-i\varphi} \right) \right\rangle$; the average being over the random phase $\varphi$. Denoting

$$X \equiv \left\langle e^{\pm i\phi} \right\rangle \tag{37}$$

we get

$$\rho_{11} = |A|^2 + |B|^2 + 2AX \, \text{Re}(B), \tag{38}$$

where $X = \left\langle e^{+i\varphi} \right\rangle = \left\langle e^{-i\varphi} \right\rangle$. Similarly,

$$\rho_{12} = \left( |A|^2 + |B|^2 \right) \cos(ka) + 2AX \, \left( \text{Re}(B) \cos(ka) + \text{Im}(B) \sin(ka) \right) \\ - i \left( |A|^2 - |B|^2 \right) \sin(ka), \tag{39}$$

and

$$\rho_{22} = |A|^2 + |B|^2 + 2AX \left( \text{Re}(B) \cos(2ka) + \text{Im}(B) \sin(2ka) \right). \tag{40}$$

Using (cf Eq. (35)), $\rho_{12} = \left( \text{Re}(K_{12}) + i \, \text{Im}(K_{12}) \right) \rho_{11}$ and $\rho_{22} = K_{22} \, \rho_{11}$, Eqs (38)-(40) constitute a set of four equations (including the real and imaginary parts of (39)) that connect between the variables $A$, $\text{Re}(B)$, $\text{Im}(B)$, $X$ and $\rho_{11}$, and can be used to express the last four in terms of $A$. Together with (35) this makes it possible to express all density matrix elements in terms of $A$ – see Appendix B for more details. The transmission and reflection coefficients, $T(E)$ and $R(E)$, respectively, are then given by

$$T(E) = \frac{\rho_{44}}{A^2} \; ; \quad R(E) = \frac{|B|^2}{A^2} \tag{41}$$

For the scattering problem depicted in Fig. 2, the treatment is similar. The incident Bloch wave of energy $E$ in the D wire is characterized by an amplitude $A$ at the "driving site" 1. The scattering center now comprises sites 3 and 4 on the D wire and sites 7 and 8 on the A wire. Scattering from this center leads to outgoing (transmitted and reflected) waves on the D and A wires, with the reflected wave on site 1 again denoted $Be^{i\varphi}$ with a random phase $\varphi$ associated with dephasing interactions in the scattering region. The equation analogous to (33) again connects all $\rho_{j1}$ elements to $\rho_{11}$ while that analogous to (34) connects all $\rho_{jk}$; $(j,k \neq 1)$ to $\rho_{jk}$; $(j \text{ or } k =1)$. The latter equation incorporates the self energies $\Sigma_A(E)$ and $\Sigma_D(E)$ at the end sites on the A and D wires, e.g. sites 5, 6 and 9 in Fig. 2, to account for the effect of the rest of the infinite chains on the dynamics of the subsystem under consideration. As above, we assume that dynamics leading to dephasing, Eq. (30) takes place only in the scattering region, i.e. $\gamma_j = \gamma \neq 0$ only for $j = 3, 4, 7, 8$. The steady states equations analogous to (33), (34) again yield Eq. (35) for all density matrix elements, and a procedure identical to that outlined above yields $\rho_{11}$ (hence all density matrix elements), $\text{Re}(B)$, $\text{Im}(B)$ and $X = \langle e^{\pm i\varphi} \rangle$ in terms of the incident amplitude $A$. The incident current is

$$J_D^{in} = \frac{2|\beta_D \sin(ka)|}{\hbar}|A|^2 = \frac{\Gamma_D(E)}{\hbar}|A|^2 \tag{42}$$

and the outgoing currents in the four channels (a, b, c, d in Fig. 2) are given by

$$J_D^{left} = \frac{2|\beta_D \sin(ka)|}{\hbar}|B|^2 = \frac{\Gamma_D(E)}{\hbar}|B|^2, \tag{43}$$

$$J_D^{right} = \frac{2\beta_D}{\hbar}\text{Im}(\overline{C}_5^* \overline{C}_4) = \frac{\Gamma_D(E)}{\hbar}\rho_{55}, \tag{44}$$

$$J_A^{left} = \frac{2\beta_A}{\hbar}\text{Im}(\overline{C}_7^* \overline{C}_8) = \frac{\Gamma_A(E)}{\hbar}\rho_{66}, \tag{45}$$

$$J_A^{right} = \frac{2\beta_A}{\hbar}\text{Im}(\overline{C}_9^* \overline{C}_8) = \frac{\Gamma_A(E)}{\hbar}\rho_{99}. \tag{46}$$

These fluxes satisfy the conservation condition

$$J_D^{in} = J_D^{left} + J_D^{right} + J_A^{left} + J_A^{right} \tag{47}$$

and can be used to obtain the current asymmetry factors $\mathcal{A}$, Eq. (28), or

$$\overline{\mathcal{A}} = \frac{J_A^{right} - J_A^{left}}{J_D^{in}}. \tag{48}$$

Figures 9-12 depict some results that show the effect of dephasing. Unless otherwise stated we set, in these calculations, the on-site energies $\alpha$ to zero, and take a nearest-neighbor coupling $\beta = 0.1\text{eV}$. The dephasing rate is varied in the range 0 … 0.5eV and the injected energy is taken within the energy band, $-0.20 \text{ eV} \leq E \leq 0.20 \text{ eV}$.

Fig. 9 shows the transmission coefficient plotted against the incident energy $E$ for the impurity scattering problem of Fig. 3 under different dephasing conditions. The impurity site, when present, is assigned site energy $\varepsilon_3 = 0.1\text{eV}$. We find that when $|\varepsilon_3| < 0.2$ (i.e. in the band) transmission decreases when dephasing increases, following the same qualitative behavior as with increasing the impurity energy. However, for the impurity energy level lying outside the energy band, $T(E)$ increases with increase in the dephasing rate. The relationship $T(E) + R(E) = 1$ is satisfied throughout. The same trends are seen also in Fig. 10, which shows the transmission coefficient (for $\alpha = 0$, $\beta = 0.1$ eV) as a function of the dephasing rate $\gamma$.

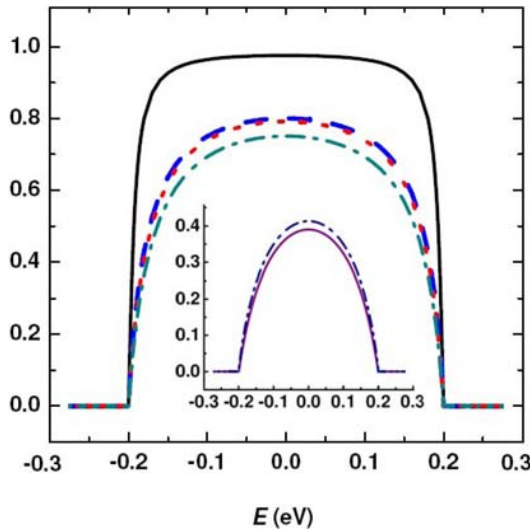

Fig. 9. Transmission $T(E)$ coefficient as a function of electron energy $E$ in the energy band $-0.20 \ eV \leq \text{E} \leq 0.20 \ eV$, for different values of dephasing on site 3. For $\varepsilon_3 = \gamma = 0$ $T(E) = 1$ (when $E$ is in the band). Other cases shown are $(\varepsilon_3, \gamma)$=(0,0.01), (0.1,0), (0.1,0.01), (0.1,0.05), shown by full (black), dashed (blue), dotted (red) and dashed-dotted (dark green) lines, respectively. The inset shows the cases (0.25,0), (0.25,0.05) in full (purple) and dash-dotted (dark blue) lines.

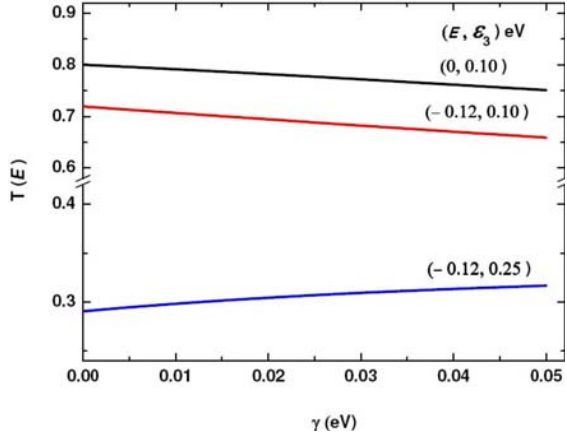

Fig. 10. Variation of transmission coefficient T($E$) with the dephasing rate $\gamma$, for different values of the incident energy $E$ (0, -0.12 eV) within the energy band and for different impurity energy levels ($\varepsilon_3 = 0.10$ eV, $0.25$ eV) that lie within or outside the band.

Absorbing boundary conditions imposed by a suitable choice of imaginary potential are often employed to facilitate numerical calculations of scattering processes. Often a constant, energy independent complex potential function is used. The self energy used in the present calculation plays the role of an energy dependent complex potential. The results of Fig. 11 shows that a proper accounting for the energy dependence of the self energy may be important: The dependence of the calculated transmission coefficient on γ shows strong dependence on Σ and an improper choice may lead to qualitatively wrong results.

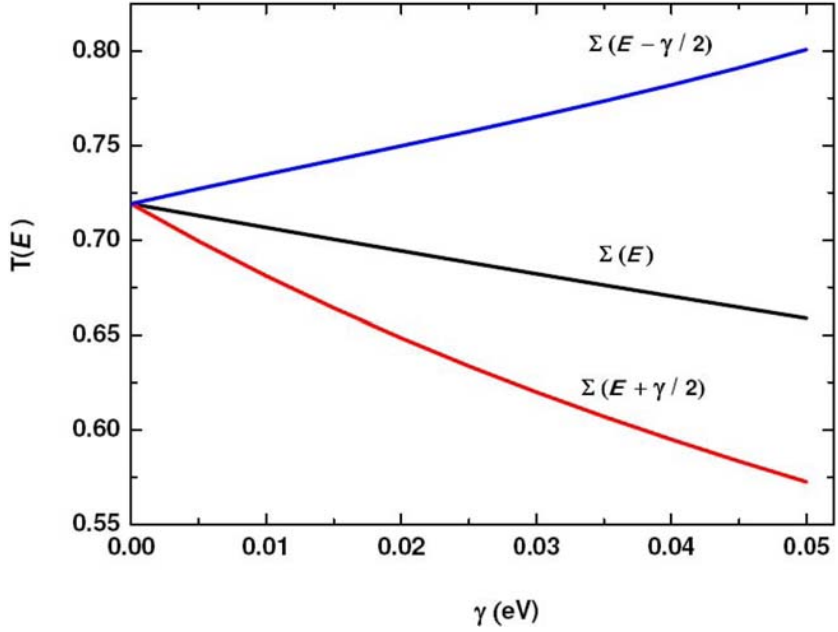

Fig. 11. Transmission coefficient *T*(*E*) plotted against the dephasing rate $\gamma$ with self energy calculated as $\Sigma(E)$ and $\Sigma(E \pm \gamma/2)$, at electron energy *E* = –0.12 eV, intersite coupling $\beta = 0.1$ eV and impurity energy 0.10 eV.

Next consider the scattering problem portrayed in Fig. 2. In what follows we use $\beta_D = \beta_A = 0.10$ eV for the intra-wire nearest-neighbor coupling, while the inter-wire coupling between sites (3, 7) and (4, 8) is taken *V* = 0.01 eV. For these model parameters and for an injection energy *E* = –0.12 eV, Fig. 12 shows the current asymmetry factor $\mathcal{A}$ (Eq. (28)) displayed as a function of the dephasing rate $\gamma$. The result obtained for the equivalent current transfer problem of Fig. 1 (Ref. [4]) for the same model parameters is seen to be almost identical. Fig. 13 shows similar results for the asymmetry factor $\bar{\mathcal{A}}$ defined by Eq. (48). Interestingly, these results depend only weakly on the imposed absorbing boundary conditions expressed by the choice of the self energy parameter $\Sigma$. We conclude, as already indicated above (Figs. 5,7) that for relatively weak interwire coupling, the current transfer model provides a good approximation to the full scattering calculation.

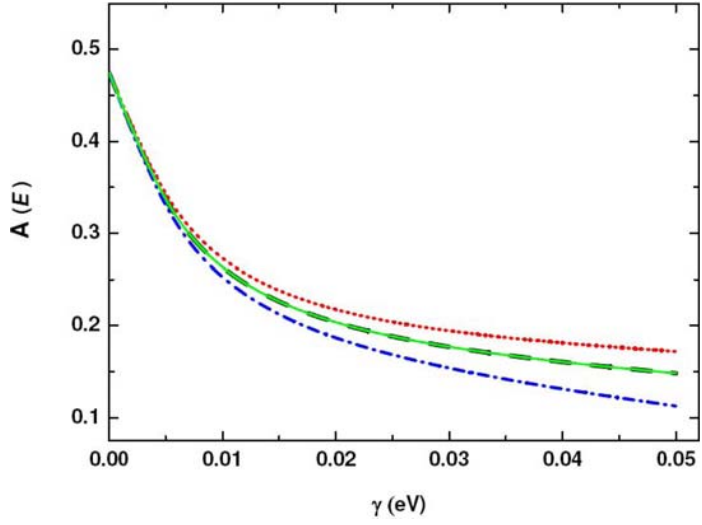

Fig. 12. Current asymmetry factor $A_1$ as a function of dephasing $\gamma$ (dashed line, black) for the coupled DA system (Fig. 2.), calculated at an electron energy $E$ = –0.12 eV with $V$ = 0.01 eV, $\beta_D = \beta_A = 0.10\ eV$, and $N_{DA}$ = 2. Also shown are results for self energy calculated as $\Sigma(E+\gamma/2)$ (dotted line, red) and $\Sigma(E-\gamma/2)$ (dashed-dotted line, blue). The full (green) line that lies on top of the dashed (black) line corresponds to the model of Fig. 1 with the same parameters, calculated with $\Sigma(E)$.

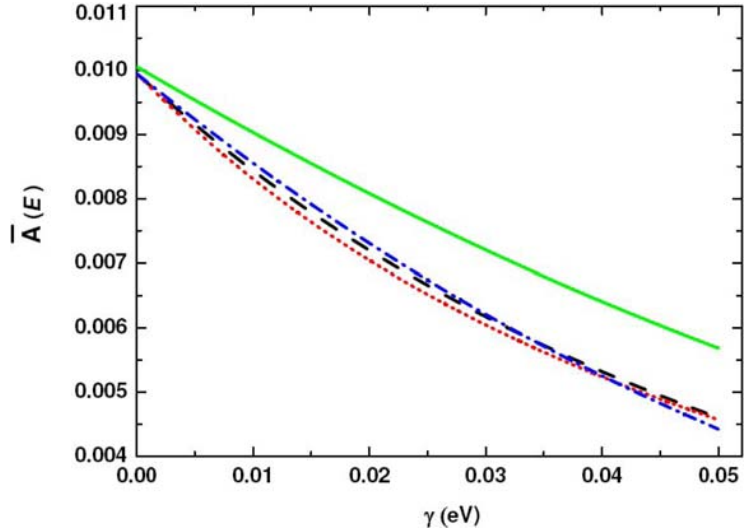

Fig. 13. Same as Fig. 12, now showing the current asymmetry factor $\bar{\mathcal{A}}$. Line designations and parameters are same as in Fig. 12.

## 4. Concluding remarks

We have compared results for transport in coupled wires systems obtained from two models that differ in boundary conditions: In the current transfer model,[4] relevant to driven systems,[1] the driving ("donor") wire is assumed to carry a given current (characterized by a Bloch wavefunction for non-interacting carriers), and the current induced in the other ("acceptor") wire is evaluated. The second model constitutes the standard scattering problem. Both models were studied in the presence of dephasing imposed in the interaction regions. We find that for weak interwire interactions the current transfer model can provide a good approximation for the full scattering problem, except near singular points associated with the band edges. In both calculations, transfer of directionality information between wires results from interference between different transfer pathways. Dephasing in the interaction region reduces the efficiency of this process, however this current transfer phenomenon is found to persist even under fairly strong dephasing.

It is interesting to note the way by which standard scattering is affected by dephasing on the target, as revealed by the present calculation. Standard scattering theory at the simplest potential scattering level can be described by an amplitude formalism, i.e. the Schrödinger equation. Scattering theory in Liouville space can describe the effect of dephasing on the target, which acts as a scattering center even in the absence of potential scattering (see dashed line (red) in Fig. 9). Such generalized scattering theory is common in describing optical scattering problems,[6-7] but is not usually used in particle scattering, where, in the context of junction transport, alternative frameworks such as the Buttiker probe model[8-9] are used. The present formalism provides a rigorous alternative that can reveal interesting physics. For example, we have found that for an impurity energy level lying inside the energy band $(\alpha - 2|\beta| \leq E \leq \alpha + 2|\beta|)$, the transmission coefficient diminishes with increasing dephasing rate, while the effect is reversed for an impurity energy level outside the band.

As discussed in Ref. [4], while more rigorous approaches (e.g. the Redfield equation[10-11]) are available, our model introduces dephasing phenomenologically, within an elastic scattering calculation. In particular, we have introduced damping of non-diagonal density matrix elements in the "site basis" and not in the eigenstates basis, which does not correspond to pure dephasing and would lead to a small inelastic scattering component ($\delta E \sim \gamma$) in the outgoing flux. This should not constitute a severe problem at dephasing rates (arising from electron-thermal phonon interactions) normally observed, but may lead to increasing errors for large dephasing rates.

**Acknowledgements**

This paper is dedicated to Prof. Eli Pollak, a friend, scientist and a leader of our field. VbM thanks the Israel Minisry of Science for a fellowship received under the program for Progressing Women in Science. The research of AN is supported by the European Science Council (FP7 /ERC grant no. 226628), the German-Israel Foundation, the Israel – Niedersachsen Research Fund, the US-Israel binational Science Foundation and the Israel Science Foundation. The research of S.S.S is supported by the University of Cyprus.

**Appendix A**

Using (13) and (6) we find

$$J_{1\to 2} = \frac{2\beta}{\hbar}\left|\bar{C}_1\right|^2 \text{Im}\left(K_2^*\right)$$
$$J_{2\to 3} = \frac{2\beta}{\hbar}\left|\bar{C}_1\right|^2 \text{Im}\left(K_3^* K_2\right) \tag{49}$$

Flux is conserved provided that

$$\text{Im}\left(K_2^*\right) = \text{Im}\left(K_3^* K_2\right) \tag{50}$$

To show this note that (13) and (14) imply

$$K_2 = \frac{1}{\beta}\left(E - \varepsilon_3 - \Sigma\right) K_3, \tag{51}$$

i.e

$$\text{Im}\left(K_3^* K_2\right) = \frac{|K_3|^2}{\beta}\text{Im}\left(E - \varepsilon_3 - \Sigma\right) = \frac{|K_3|^2}{2\beta}\Gamma \tag{52}$$

(50) therefore holds if

$$\frac{\text{Im}\left(K_2^*\right)}{|K_3|^2} = \text{Im}\frac{K_2^*}{|K_3|^2} = \text{Im}\frac{1}{K_3}\left(\frac{K_2}{K_3}\right)^* = \frac{\Gamma}{2\beta} \tag{53}$$

Indeed, from (51) and (13) we get

$$\frac{1}{K_3}\left(\frac{K_2}{K_3}\right)^* = \frac{|E-\varepsilon_3-\Sigma|^2}{\beta^2}\frac{1}{K_2} \tag{54}$$

while (13) itself implies

$$\frac{1}{K_2} = \frac{1}{\beta}\left(E-E_1-\frac{\beta^2}{E-\varepsilon_3-\Sigma}\right) \Rightarrow \operatorname{Im}\frac{1}{K_2} = \frac{\beta\Gamma}{2|E-\varepsilon_3-\Sigma|^2} \tag{55}$$

Using (54) and (55) it is easy to show that (53) holds.

**Appendix B**

Here provide the explicit results for $\rho_{11}$, $\operatorname{Re}(B)$, $\operatorname{Im}(B)$ and $X$ in terms of the incident amplitude A. Using (from (35)) $\operatorname{Re}(\rho_{12}) = \operatorname{Re}(K_{12})\rho_{11}$, $\operatorname{Im}(\rho_{12}) = \operatorname{Im}(K_{12})\rho_{11}$ and $\rho_{22} = K_{22}\rho_{11}$ as well as Eqs. (38) - (40) we obtain

$$|B|^2 = \left(\frac{2A\operatorname{Im}(K_{12})}{\sin(ka)-\operatorname{Im}(K_{12})}\right)X\operatorname{Re}(B) + \left(\frac{\sin(ka)+\operatorname{Im}(K_{12})}{\sin(ka)-\operatorname{Im}(K_{12})}\right)|A|^2, \tag{56}$$

$$X\operatorname{Re}(B) = \frac{(R_4 - R_1 X\operatorname{Im}(B))}{R_5}, \tag{57}$$

$$X\operatorname{Im}(B) = \frac{(R_3 R_5 - R_4 R_6)}{(R_2 R_5 - R_1 R_6)}, \tag{58}$$

where

$$R_1 = 2A\sin(ka),$$
$$R_2 = 2A\sin(2ka),$$
$$R_3 = -(1-K_{22})\left[1+\frac{\sin(ka)+\operatorname{Im}(K_{12})}{\sin(ka)-\operatorname{Im}(K_{12})}\right]|A|^2,$$
$$R_4 = -(\cos(ka)-\operatorname{Re}(K_{12}))\left[1+\frac{\sin(ka)+\operatorname{Im}(K_{12})}{\sin(ka)-\operatorname{Im}(K_{12})}\right]|A|^2,$$
$$R_5 = 2(\cos(ka)-\operatorname{Re}(K_{12}))\left[1+\frac{\operatorname{Im}(K_{12})}{\sin(ka)-\operatorname{Im}(K_{12})}\right]A,$$
$$R_6 = 2\left[(1-K_{22})\frac{\operatorname{Im}(K_{12})}{\sin(ka)-\operatorname{Im}(K_{12})}+(\cos(2ka)-K_{22})\right]A.$$